\documentclass[twocolumn]{aastex631}
\usepackage{amsmath}

\usepackage[T1]{fontenc}

\shorttitle{ALMA observations of PSR B1259-63}
\shortauthors{Fujita et al.}

\begin{document}

\title{ALMA observations of the gamma-ray binary system PSR B1259-63/LS 2883 during the 2024 periastron passage}
\author[0000-0003-0058-9719]{Yutaka Fujita}
\affiliation{Department of Physics, Graduate School of Science
Tokyo Metropolitan University\\
1-1 Minami-Osawa, Hachioji-shi, Tokyo 192-0397, Japan}

\author{Akiko Kawachi}
\affiliation{Department of Physics, School of Science, Tokai University, Kitakaname, Hiratsuka, Kanagawa 259-1292,
Japan}

\author{Atsuo T. Okazaki}
\affiliation{Center for Development Policy Studies, Hokkai-Gakuen University, 1-40, 4-chome, Asahi-machi, Toyohira-ku, Sapporo,
Hokkaido 062-8605, Japan}

\author{Hiroshi Nagai}
\affiliation{ALMA Project, National Astronomical Observatory Japan, 2-21-1 Osawa, Mitaka, Tokyo 181-8588, Japan}
\affiliation{The Graduate University for Advanced Studies, SOKENDAI, 2-21-1 Osawa, Mitaka, Tokyo 181-8588, Japan}

\author{Norita Kawanaka}
\affiliation{Department of Physics, Graduate School of Science
Tokyo Metropolitan University\\
1-1 Minami-Osawa, Hachioji-shi, Tokyo 192-0397, Japan}
\affiliation{ALMA Project, National Astronomical Observatory Japan, 2-21-1 Osawa, Mitaka, Tokyo 181-8588, Japan}

\author{Takuya Akahori}
\affiliation{Mizusawa VLBI Observatory, National Astronomical Observatory Japan, 2-21-1 Osawa, Mitaka, Tokyo 181-8588, Japan}

\begin{abstract}

We present observations of the gamma-ray binary PSR B1259-63/LS 2883 with the Atacama Large Millimeter/submillimeter Array (ALMA) at Bands 3 (97 GHz), 6 (233 GHz), and 7 (343 GHz). PSR B1259-63/LS 2883 consists of a pulsar in a highly eccentric orbit around a massive companion star, with the pulsar passing through the circumstellar disk near periastron. Our new data were obtained over several epochs, ranging from -61 to +29 days from the periastron passage in 2024. We report an increase in flux in all bands near the periastron. The significant change in Band 3 flux suggests synchrotron emission from the interaction between the pulsar wind and the stellar wind or disk. The Band 6 flux shows an increase around periastron and a transition from thermal emission from the circumstellar disk to synchrotron emission. The Band 7 observation +24 days after periastron shows a brightening, suggesting that the pulsar's passage through the disk does not result in its immediate destruction. We discuss the implications of these results for the interaction between the pulsar wind and the circumstellar disk, such as the possible disk expansion after periastron.

\end{abstract}


\section{Introduction} \label{sec:intro}

Gamma-ray binaries are a rare and intriguing class of high-energy astrophysical systems characterized by strong gamma-ray emission \citep{2013A&ARv..21...64D}. Among these, PSR B1259-63/LS~2883 (hereafter B1259) stands out as a particularly interesting object, consisting of the pulsar PSR B1259-63 in an eccentric orbit around a massive late Oe-type companion star (LS~2883) with an equatorial decretion disk \citep{2011ApJ...732L..11N}. B1259 is located at a distance of $\sim 2.6$~kpc from Earth \citep{2018MNRAS.479.4849M} and has an orbital period of about 3.4 years, with an eccentricity of 0.87 \citep{2014MNRAS.437.3255S}.

This system provides a unique laboratory for studying the interactions between pulsar winds and stellar environments, as well as the processes governing the acceleration and emission of high-energy particles in binary systems.
The most dramatic events in this binary occur near periastron, when the pulsar PSR B1259-63 passes through or near the circumstellar disk of the companion star LS~2883. During and near these passages, complex interactions between the pulsar wind and the stellar environment lead to enhanced high-energy emission across the electromagnetic spectrum, from radio to gamma rays (e.g. \citealt{2021Univ....7..242C}, and references therein).
No pulsed flux was detected within $t_p\sim \pm 15$~days\footnote{We refer to the day of each periastron passage as $t_p=0$~days.} \citep{1996MNRAS.279.1026J}. The eclipse of the pulsar during this period is likely due to absorption and severe pulse scattering by the companion star's disk \citep{1996MNRAS.279.1026J}. This eclipse of the pulsed emission is accompanied by an increase in the unpulsed radio flux, which begins at $t_p\sim -30$~days and reaches a maximum at $t_p\sim -10$~days. After the periastron passage, the flux decreases before climbing to a second peak at $t_p\sim +20$~days \citep{1999MNRAS.302..277J,2005MNRAS.358.1069J}.
We note that the radio light curve is highly variable from cycle to cycle. For example, the timing of the flux peaks and the rate of flux decline after the second peak change at each periastron passage (Figure~1 in \citet{2021Univ....7..242C}).

Previous studies have extensively investigated the behavior of B1259 at various wavelengths, particularly focusing on the low-frequency ($\lesssim 10$~GHz) radio, X-ray and gamma-ray regimes \citep[e.g.][]{1996MNRAS.279.1026J,2005A&A...442....1A,2006MNRAS.367.1201C,2009ApJ...698..911U,2011ApJ...736L..11A,2021Univ....7..472C,2024MNRAS.528.5231C,2024A&A...687A.219H}. However, observations at millimeter and submillimeter wavelengths ($\sim 100$--300~GHz) have been comparatively limited.

\begin{deluxetable*}{cccccccccccc}
\tablecaption{Summary of ALMA observations\label{tab:sum}}
\tablewidth{700pt}
\tabletypesize{\scriptsize}
\tablehead{Date & Day & Band & Freq.$^{\rm a}$ & $N_{\rm ant}^{\rm b}$ & $T_{\rm on}^{\rm c}$ & Bandpass/flux$^{\rm d}$ & Gain$^{\rm e}$ & Beam shape & PA$^{\rm f}$ & Image rms & Observed flux \\
\colhead{} & (from $t_p$) &  & (GHz) & \colhead{} & (min) & \colhead{} & \colhead{} & (arcsec) & (deg) & ($\rm\mu Jy\: bm^{-1}$) & (mJy)
} 
\startdata
30-Apr & -61 & 3 & 97 & 44 & 6 & J1427-4206 & J1308-6707 & $2.0\times 1.7$ & 61 & 29 & $0.32\pm 0.06$ \\
30-Apr & -61 & 6 & 233 & 44 & 5 & J1427-4206 & J1254-6111 & $0.86\times 0.81$ & 42 & 68 & $0.93\pm 0.18$ \\
4-May & -57 & 7 & 343 & 42 & 5 & J1617-5848 & J1308-6707 & $0.57\times 0.50$ & 46 & 122 & $1.60\pm 0.27$ \\
30-May & -31 & 6 & 233 & 45 & 5 & J1427-4206 & J1308-6707 & $0.41\times 0.30$ & -16 & 38 & $0.72\times 0.08$ \\
2-Jun & -28 & 3 & 97 & 41 & 5 & J1617-5848 & J1308-6707 & $0.92\times 0.54$ & 43 & 32 & $0.26\pm 0.06$ \\
29-Jun & -1 & 3 & 97 & 45 & 6 & J1617-5848 & J1308-6707 & $0.48\times 0.39$ & -31 & 39 & $3.17\pm 0.20$ \\
29-Jun & -1 & 6 & 233 & 45 & 5 & J1617-5848 & J1308-6707 & $0.18\times 0.16$ & -19 & 48 & $2.48\pm 0.23$ \\
24-Jul & +24 & 7 & 343 & 41 & 5 & J1617-5848 & J1308-6707 & $0.16\times 0.11$ & -11 & 130 & $3.11\pm 0.54$ \\
29-Jul & +29 & 3 & 97 & 42 & 6 & J1427-4206 & J1308-6707 & $0.78\times 0.59$ & -23 & 31 & $1.34\pm 0.10$ \\
29-Jul & +29 & 6 & 233 & 42 & 5 & J1427-4206 & J1308-6707 & $0.35\times 0.27$ & -31 & 44 & $1.89\pm 0.15$ \\
\enddata
\tablecomments{$^{\rm a}$Center frequency, $^{\rm b}$Number of antennas used for observation, $^{\rm c}$Total integration time of the target source, $^{\rm d}$Band pass and flux calibrator name, $^{\rm e}$Gain calibrator name, and $^{\rm f}$Beam position angle.
}
\end{deluxetable*}

The advent of the Atacama Large Millimeter/submillimeter Array (ALMA) allowed us to study the emission at higher frequencies ($\gtrsim 100$~GHz) with excellent sensitivity. In 2017, we observed the binary in Band~3 (97 GHz) and Band~7 (343 GHz) from $t_p=+69$ to +84~days (\citealt{2019PASJ...71L...3F}, hereafter Paper~I). 
We found that the Band 3 flux is identical to an extrapolation of the unpulsed fluxes at lower frequencies, suggesting that the Band 3 emission is synchrotron, as is the low frequency emission. The emission is likely produced by electrons accelerated by pulsar wind shocks generated by the interaction between the pulsar wind and the circumstellar material of the companion star \citep{2012ApJ...750...70T}. In contrast to the Band~3 flux, the Band~7 flux at $t_p=+69$~days is much larger than the extrapolated value. The most plausible explanation is that this emission is thermal (blackbody and/or bremsstrahlung) and comes from the circumstellar disk (Paper~I). This disk evidence provides an important clue to the origin of the non-thermal emission from the binary, since the gamma-ray emission can be produced by inverse Compton (IC) scattering of photons from the disk \citep{2012MNRAS.426.3135V,2019A&A...627A..87C}.

In 2019, we observed the binary with ALMA when the pulsar was close to the apastron (\citealt{2020PASJ...72L...9F}, hereafter Paper~II) and found that both the Band 3 ($t_p=+771$~days) and 7 ($+776$~days) fluxes had decreased from those in 2017. The Band 3 flux is consistent with the extrapolation of pulsed fluxes at low frequencies. This means that while the unpulsed emission produced by the pulsar-companion star interaction had disappeared, the pulsed emission intrinsic to the pulsar was observed. The decrease in Band 7 flux indicates that the disk is not perfectly stationary, but is evolving even away from periastron.

In this study we present new observations of B1259 in 2024 with ALMA, covering three frequency bands. These observations were aimed at capturing for the first time the behavior of the system around the periastron passage.

\section{Observations and Data Reduction}

We performed observations of B1259 with ALMA around the 2024 periastron passage (June 30). The observations were made in Band 3 (97 GHz), Band 6 (233 GHz), and Band 7 (343 GHz). 
The ALMA observations used ranged from 41 to 45 antennas, providing beam sizes from $\sim 0.2''$ to $2''$ depending on the frequency band and array configuration (Table~\ref{tab:sum}). The total on-source integration time for each observation ranged from 5 to 6 minutes.
The data were processed using the Common Astronomy Software Applications (CASA) package version 6.5.4.9 \citep{2007ASPC..376..127M} and ALMA Pipeline version 2023.1.0.124 in a standard manner.
The bandpass and gain calibrators are summarized in Table~\ref{tab:sum}. The flux scaling was derived on the bandpass calibrator using flux information provided by the Joint ALMA Observatory (JAO). The observing conditions with respect to weather and precipitable water vapor (PWV) were normal in the three bands.

After calibration, we created images for each observation using the CLEAN algorithm within CASA. We employed Briggs weighting with a robust parameter of 0.5, balancing between sensitivity and resolution. The resulting synthesized beam sizes are presented in Table~\ref{tab:sum}. The image size of B1259 is consistent with a point source and the upper limit is given by the beam sizes.

\section{Results}

\subsection{Light Curves}

\begin{figure}[t]
\plotone{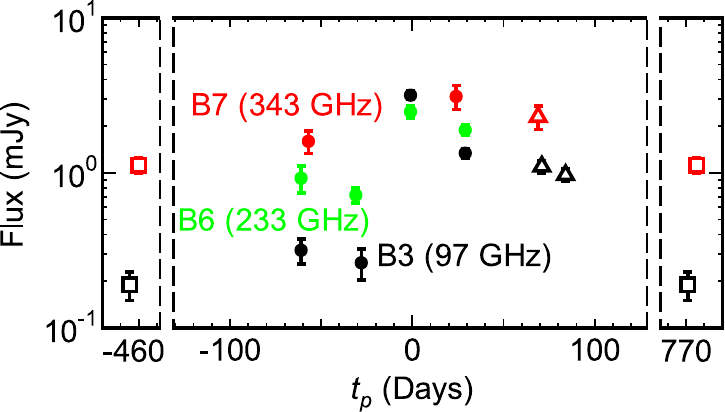}
\caption{Radio light curves of B1259. The black, light green, and red symbols represent Band~3 (97 GHz), Band~6 (233 GHz), and Band~7 (343 GHz), respectively. The solid circles represent the 2024 observations. For reference, the 2017 (Paper~I) and 2019 (Paper~II) observations are represented by open triangles and open squares, respectively. Note that the 2019 observations shown at $t_p\sim -460$~days and those shown at $t_p\sim +770$~days are identical.
\label{fig:LC}}
\end{figure}

Our ALMA observations of B1259 reveal significant variability in the millimeter and submillimeter emission across different orbital phases and frequency bands. The observed fluxes are listed in Table~\ref{tab:sum} and the light curves for the three bands are shown in Figure~\ref{fig:LC}.

In Band~3 (97~GHz), the flux remains relatively low and constant during the early approach to periastron ($t_p=-61$ and -28~days). However, we observe a dramatic increase in flux as the system approaches periastron, with the emission peaking at $3.17\pm 0.20$~mJy at $t_p= -1$~days. This is a tenfold increase from pre-periastron levels. At $t_p=+29$~days the flux begins to decline, but remains elevated compared to the initial observations, measuring $1.34\pm 0.10$~mJy.
The significant flux increase at $t_p\gtrsim -20$~days is similar to that at lower frequencies ($\lesssim 10$~GHz; \citealt{2005MNRAS.358.1069J}), suggesting synchrotron emission from relativistic electrons \citep{1999ApJ...514L..39B,2002MNRAS.336.1201C}.
We note that since the number of our ALMA observations is small, the double peaks found in the light curves for the low frequency radio and X-ray bands \citep{2021Univ....7..242C}, associated with the disk passage of the pulsar at $t_p\sim -10$ and $+20$~days, cannot be discussed for our observations.

The emission at Band~6 (233~GHz) shows a general increase as the system approaches periastron, similar to the trend observed in Band~3 (Figure~\ref{fig:LC}). However, the relative amplitude of this increase is less pronounced, with the flux near periastron ($2.48 \pm 0.23$~mJy) being only about 2.7 times higher than the initial measurement ($0.93 \pm 0.18$~mJy at $t_p=-61$~days).
Interestingly, the flux at $t_p=-1$~day from periastron ($2.48 \pm 0.23$~mJy) is lower than the corresponding measurement in Band~3 ($3.17 \pm 0.20$~mJy). This inversion of the spectrum suggests a transition in the dominant emission mechanism, from the thermal to the synchrotron emission, around these two frequencies (see Section~\ref{sec:SED}).
The post-periastron measurement at $t_p=+29$~days ($1.89 \pm 0.15$~mJy) shows a slower decay compared to Band 3, remaining at about 76\% of the near-periastron value. This behavior hints at a possible contribution from thermal emission from the companion star's circumstellar disk, which would be expected to vary more slowly than the synchrotron component.

Band 7 observations at 343 GHz provide insights into the system's higher frequency behavior, despite fewer data points. The pre-periastron measurement at $t_p =-57$ days is higher than the corresponding measurements in Bands 3 and 6. This suggests a positive spectral index ($\alpha>0$), where the index is defined as $S_\nu \propto \nu^{\alpha}$, in the undisturbed state before the pulsar passes the disk.
The post-periastron observation at $t_p =+24$ days shows a slight increase in flux to $3.11 \pm 0.54$~mJy. 

For reference, we show the results of our 2017 (Paper~I) and 2019 (Paper~II) observations in Figure~\ref{fig:LC}, although the flux at a given $t_p$ can vary for each orbital period. While the flux in Band~3 varies significantly during an orbital period, the flux in Band~7 does not,

\begin{figure}[ht]
\plotone{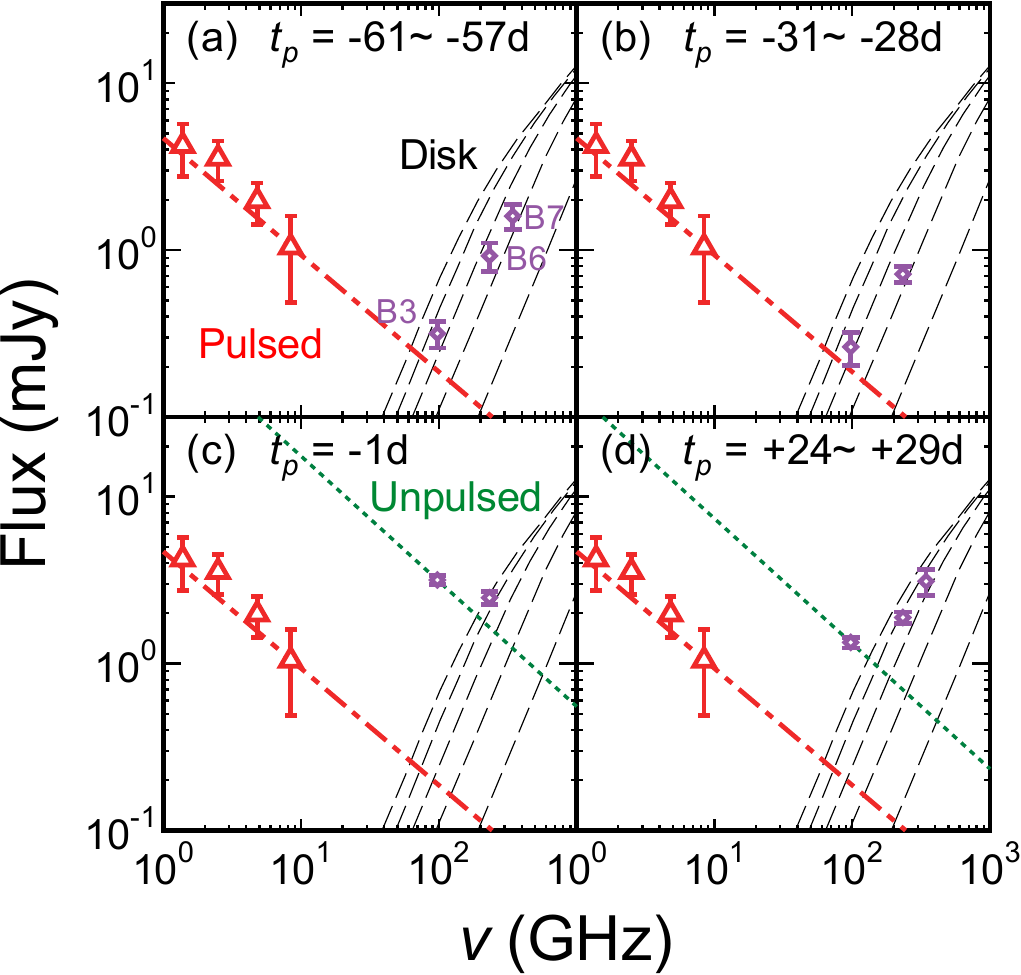}
\caption{Radio SEDs of B1259. Our new ALMA observations in 2024 are indicated by open purple diamonds. (a) From $t_p=-61$~days to -57~days. (b) From $t_p=-31$~days to -28~days. (c) At $t_p=-1$~days. (d) From $t_p=+24$~days to +29~days. The exact observation dates are shown in the Table~\ref{tab:sum}. 
Averaged pulsed fluxes obtained with ATCA before and after the 2004 passage are shown by red open triangles \citep{2005MNRAS.358.1069J}. 
The associated red bars are the standard deviations of the data at each frequency. The dash–dotted red line is a spectrum represented by $S_\nu\propto \nu^{-0.7}$. The normalization is set so that the line passes through the 2019 ALMA observation at 97~GHz (see Figure~3 in Paper~II). 
The dotted green line is a typical unpulsed spectrum for $+10\lesssim t_p\lesssim +30$~days for the 2021 periastron passage, represented by $S_\nu\propto \nu^{-0.75}$ \citep{2024MNRAS.528.5231C}. The normalization is set so that the line passes through the ALMA observations at 97~GHz in (c) and (d), respectively. 
The dashed black lines show the infrared emission from the circumstellar disk of LS~2883 for different disk sizes predicted by \citet{2011MNRAS.412.1721V}. The disk sizes are (from bottom to top) 10, 20, 30, 40, and 50~$R_\star$.
\label{fig:spec}}
\end{figure}

\subsection{Spectral Energy Distributions}
\label{sec:SED}

To better understand the emission mechanisms at play, we constructed spectral energy distributions (SEDs) for the system at different orbital phases. Figure~\ref{fig:spec} shows the SEDs at four key phases: early approach ($t_p =-61$~to -57 days), mid-approach (-31 to -28 days), near-periastron (-1 day), and post-periastron (+24 to +29 days).

When the pulsar is approaching periastron (Figures~\ref{fig:spec}(a) and (b)), the spectrum rises with frequency. The fluxes are close to those at $t_p\sim +770$~days (Figure~\ref{fig:LC}), suggesting that the Band~3 emission is the pulsed emission from the pulsar (dash-dotted red line in Figure~\ref{fig:spec}), while the Band~6 and 7 emissions are the thermal radiation from the circumstellar disk (dashed black lines in Figure~\ref{fig:spec}; see Paper~II).

Near periastron (Figure~\ref{fig:spec}(c)), the emission, especially in the lower frequency (Band~3), increases dramatically, causing an inversion in the spectrum between Bands~3 and~6. This suggests the emergence of unpulsed synchrotron emission from electrons accelerated as the pulsar passes through the stellar wind and/or circumstellar disk (Papers~I and~II). The synchrotron emission appears to outshine the thermal emission at Band~6.
However, the spectral index between Bands~3 and~6 is $\alpha\sim -0.3$. The slope is less steep compared to a typical unpulsed synchrotron emission ($\alpha\sim -0.75$; dotted green line in Figure~\ref{fig:spec}), which could indicate a contribution of the disk emission to the Band~6 emission.

After periastron (Figure~\ref{fig:spec}(d)), the spectrum rises again with frequency, although the absolute fluxes are larger than before periastron (Figures~\ref{fig:spec}(a) and (b)). The emission seems to come from the disk again, especially in Bands~6 and 7. However, the spectrum between Bands~3 and 6 is less steep than that between Bands~6 and~7, suggesting that the unpulsed synchrotron emission may contribute to the Band~3 emission (dotted green line).

\section{Discussion}

Our multi-band ALMA observations of B1259 reveal a complex and dynamic picture of the binary system's millimeter and submillimeter emission. 

The dramatic increase in the Band 3 flux near periastron (Figure~\ref{fig:LC}) is strongly suggestive of synchrotron emission from relativistic electrons. This is because the behavior is consistent with observations at lower radio frequencies \citep{2002MNRAS.336.1201C,2005MNRAS.358.1069J}.
As the pulsar approaches periastron, it encounters increasingly dense regions of the stellar wind and/or circumstellar disk. The collision between the fast pulsar wind and the slower, denser stellar material creates a shock front where particles can be accelerated to relativistic energies \citep{1997ApJ...477..439T}. 
The strength of this interaction, and consequently the intensity of the synchrotron emission, peaks near periastron when the pulsar is closest to or possibly passing through the densest part of the disk.
However, in detail the Band 3 flux at $t_p= +29$~days is smaller than that at $t_p=-1$~days, while they are comparable at lower frequencies \citep{2021Univ....7..242C}. Moreover, the X-ray emission, which is also considered to be synchrotron radiation, is much stronger at $t_p\sim +29$~days than at periastron \citep{2021Univ....7..242C}. This suggests that there are multiple electron components such as weakly and strongly shocked electrons \citep{2020MNRAS.497..648C}. The clumpiness of the stellar wind may also affect the synchrotron radiation \citep{2020MNRAS.497..648C}, although the details are beyond the scope of this study.

While the Band 3 observations are dominated by synchrotron emission, the positive slope $\alpha>0$ spectra at higher frequencies (Bands~6 and~7 in Figures~\ref{fig:spec}(a), (b), and (d)) reveal a significant contribution from thermal emission from the companion's circumstellar disk. The persistence and even enhancement of the emission at Band~7 after periastron passage (Figure~\ref{fig:LC}) suggests that the circumstellar disk remains largely intact despite the pulsar's close approach. This finding is consistent with hydrodynamical simulations, which predict that the disk may be perturbed but not completely disrupted by the pulsar's passage \citep{2011PASJ...63..893O}. In fact, a comparison with the thermal emission model of \citet{2011MNRAS.412.1721V} shows that the disk size is $\sim 20$--$30\: R_\star$, where $R_\star$ is the radius of the companion, well before periastron (Figure~\ref{fig:spec}(a)), while it is $\sim 40\: R_\star$ at $t_p\sim +24$--29~days (Figure~\ref{fig:spec}(d)). Note that the latter size is comparable to the binary separation at that epoch.

The observed increase in Band 7 flux after periastron could be due to disk expansion. 
\citet{2021PASJ...73..545K} observed the variability of B1259 in the near infrared (NIR) bands with the IRSF (InfraRed Survey Facility) 1.4 m telescope. The observations were mainly performed for about a month around the periastron passages of 2010 and 2014.
The NIR luminosity change in \citet{2021PASJ...73..545K} is thought to be the result of the disk being stretched by the tidal interaction, but the resulting brightening is only $\lesssim 0.1$ mag ($\lesssim 10$\%). In contrast, the change in luminosity at Band 7 was about 200\%.
The radio emission is emitted from a more outer region of the disk than the NIR emission (e.g. \citealt{2006ApJ...639.1081C}; see also \citealt{2013A&ARv..21...69R}) and is therefore more subject to tidal interactions. However, even taking this into account, it would be difficult to explain the 2-fold brightening by tidal interactions alone. In view of these considerations, the brightening in Band 7 may be due to the effect of pulsar winds pushing the disk outward \citep{2011PASJ...63..893O,2012ApJ...750...70T}.

Previous observations have shown that the H$\alpha$ equivalent width of B1259 also has a large ($\sim 30$\%) post-periastron increase: It peaks at $t_p\sim +10$--$+20$ days and starts to decrease at $t_p\sim +30$~days \citep{2021Univ....7..242C,2024arXiv241102128C}. This behavior is thought to be related to changes in the disk structure not directly associated with the so-called GeV flares \citep{2016MNRAS.455.3674V,2024MNRAS.528.5231C}. If future Band 7 observations with higher cadence show a similar light curve, it will support the idea that the Band 7 emission comes from the part of the disk where the H$\alpha$ line arises, but is not related to GeV flares. Conversely, if a significant difference is observed between the Band 7 light curve and variations in the H$\alpha$ equivalent width, this will allow us to trace how disturbances in the disk evolve between the radio-emitting and H$\alpha$-emitting regions. Hydrodynamical simulations with radiative transfer could provide valuable insights into the detailed evolution of the disk.

In general, the Band~7 light curve, including the 2017 and 2019 observations (Figure~\ref{fig:LC}), shows that the circumstellar disk is relatively stable. This stability is critical for understanding the high-energy phenomena observed in this system over several periastron passages.
For example, GeV flares \citep{2011ApJ...736L..11A} and X-ray ejecta \citep{2011ApJ...730....2P,2019ApJ...882...74H} could be associated with a partial destruction of the circumstellar disk. 
Indeed, GeV flares have been observed \citep{2024arXiv241102128C} when the Band~7 flux increased (Figure~\ref{fig:spec}(d)), which may be caused by the disk expansion with the partial destruction.

\section{Conclusions}

Our multi-band ALMA observations of the gamma-ray binary B1259 have provided new insights into the complex interplay between the pulsar and the circumstellar disk of the companion star. The main results of our study are:

1. Strong variability in the Band 3 (97 GHz) emission, peaking near periastron, consistent with synchrotron radiation from shock-accelerated electrons in the pulsar-stellar wind interaction.

2. A more complex behavior in Band 6 (233 GHz), suggesting a transition between synchrotron and thermal emission mechanisms: the synchrotron emission becomes dominant at periastron ($t_p\sim -1$~days), whereas the thermal emission is the dominant emission mechanism at other phases.

3.  The post-periastron brightening in Band 7 (343 GHz) indicates the resilience of the circumstellar disk and possible expansion of the disk.

These results demonstrate the power of millimeter and submillimeter observations to probe the physical conditions and processes in gamma-ray binaries. The persistence of the circumstellar disk, as evidenced by our Band 7 data, has important implications for understanding the long-term stability and nature of the high-energy phenomena in B1259.

\begin{acknowledgments}
We thank the reviewer for helpful comments that improved the paper. We also thank the EA ALMA
Regional Center (EA-ARC) for their support. This work was
supported by NAOJ ALMA Scientific Research grant Code
2022-21A, and JSPS KAKENHI grant Nos. JP22H01268,
JP22H00158, JP22K03624, JP23H04899 (Y.F.), JP21H01137, JP18K03709
(H.N.), and JP22K03686 (N.K.).
This paper makes use of the following ALMA data: ADS/JAO.ALMA\#2017.1.01188.S, 2019.1.00320.S, and 2023.1.00271.S. ALMA is a partnership of ESO (representing its member states), NSF (USA) and NINS (Japan), together with NRC (Canada), MOST and ASIAA (Taiwan), and KASI (Republic of Korea), in cooperation with the Republic of Chile. The Joint ALMA Observatory is operated by ESO, AUI/NRAO and NAOJ. Data analysis was in part carried out on the Multi-wavelength Data
Analysis System operated by the Astronomy Data Center (ADC),
National Astronomical Observatory of Japan.
\end{acknowledgments}

\bibliography{B1259}{}

\begin{thebibliography}{}
\expandafter\ifx\csname natexlab\endcsname\relax\def\natexlab#1{#1}\fi
\providecommand{\url}[1]{\href{#1}{#1}}
\providecommand{\dodoi}[1]{doi:~\href{http://doi.org/#1}{\nolinkurl{#1}}}
\providecommand{\doeprint}[1]{\href{http://ascl.net/#1}{\nolinkurl{http://ascl.net/#1}}}
\providecommand{\doarXiv}[1]{\href{https://arxiv.org/abs/#1}{\nolinkurl{https://arxiv.org/abs/#1}}}

\bibitem[{{Abdo} {et~al.}(2011){Abdo}, {Ackermann}, {Ajello}, {Allafort},
  {Ballet}, {Barbiellini}, {Bastieri}, {Bechtol}, {Bellazzini}, {Berenji},
  {Blandford}, {Bonamente}, {Borgland}, {Bregeon}, {Brigida}, {Bruel},
  {Buehler}, {Buson}, {Caliandro}, {Cameron}, {Camilo}, {Caraveo}, {Cecchi},
  {Charles}, {Chaty}, {Chekhtman}, {Chernyakova}, {Cheung}, {Chiang},
  {Ciprini}, {Claus}, {Cohen-Tanugi}, {Cominsky}, {Corbel}, {Cutini},
  {D'Ammando}, {de Angelis}, {den Hartog}, {de Palma}, {Dermer}, {Digel},
  {Silva}, {Dormody}, {Drell}, {Drlica-Wagner}, {Dubois}, {Dubus}, {Dumora},
  {Enoto}, {Espinoza}, {Favuzzi}, {Fegan}, {Ferrara}, {Focke}, {Fortin},
  {Fukazawa}, {Funk}, {Fusco}, {Gargano}, {Gasparrini}, {Gehrels}, {Germani},
  {Giglietto}, {Giommi}, {Giordano}, {Giroletti}, {Glanzman}, {Godfrey},
  {Grenier}, {Grondin}, {Grove}, {Grundstrom}, {Guiriec}, {Gwon}, {Hadasch},
  {Harding}, {Hayashida}, {Hays}, {J{\'o}hannesson}, {Johnson}, {Johnson},
  {Johnston}, {Kamae}, {Katagiri}, {Kataoka}, {Keith}, {Kerr},
  {Kn{\"o}dlseder}, {Kramer}, {Kuss}, {Lande}, {Lee}, {Lemoine-Goumard},
  {Longo}, {Loparco}, {Lovellette}, {Lubrano}, {Manchester}, {Marelli},
  {Mazziotta}, {Michelson}, {Mitthumsiri}, {Mizuno}, {Moiseev}, {Monte},
  {Monzani}, {Morselli}, {Moskalenko}, {Murgia}, {Nakamori}, {Naumann-Godo},
  {Neronov}, {Nolan}, {Norris}, {Noutsos}, {Nuss}, {Ohsugi}, {Okumura},
  {Omodei}, {Orlando}, {Paneque}, {Parent}, {Pesce-Rollins}, {Pierbattista},
  {Piron}, {Porter}, {Possenti}, {Rain{\`o}}, {Rando}, {Ray}, {Razzano},
  {Razzaque}, {Reimer}, {Reimer}, {Reposeur}, {Ritz}, {Sadrozinski}, {Scargle},
  {Sgr{\`o}}, {Shannon}, {Siskind}, {Smith}, {Spandre}, {Spinelli},
  {Strickman}, {Suson}, {Takahashi}, {Tanaka}, {Thayer}, {Thayer}, {Thompson},
  {Thorsett}, {Tibaldo}, {Tibolla}, {Torres}, {Tosti}, {Troja}, {Uchiyama},
  {Usher}, {Vandenbroucke}, {Vasileiou}, {Vianello}, {Vitale}, {Waite}, {Wang},
  {Winer}, {Wolff}, {Wood}, {Wood}, {Yang}, {Ziegler}, \&
  {Zimmer}}]{2011ApJ...736L..11A}
{Abdo}, A.~A., {Ackermann}, M., {Ajello}, M., {et~al.} 2011, \apjl, 736, L11,
  \dodoi{10.1088/2041-8205/736/1/L11}

\bibitem[{{Aharonian} {et~al.}(2005){Aharonian}, {Akhperjanian}, {Aye},
  {Bazer-Bachi}, {Beilicke}, {Benbow}, {Berge}, {Berghaus}, {Bernl{\"o}hr},
  {Boisson}, {Bolz}, {Braun}, {Breitling}, {Brown}, {Bussons Gordo},
  {Chadwick}, {Chounet}, {Cornils}, {Costamante}, {Degrange},
  {Djannati-Ata{\"\i}}, {O'C. Drury}, {Dubus}, {Emmanoulopoulos}, {Espigat},
  {Feinstein}, {Fleury}, {Fontaine}, {Fuchs}, {Funk}, {Gallant}, {Giebels},
  {Gillessen}, {Glicenstein}, {Goret}, {Hadjichristidis}, {Hauser},
  {Heinzelmann}, {Henri}, {Hermann}, {Hinton}, {Hofmann}, {Holleran}, {Horns},
  {de Jager}, {Johnston}, {Kh{\'e}lifi}, {Kirk}, {Komin}, {Konopelko},
  {Latham}, {Le Gallou}, {Lemi{\`e}re}, {Lemoine-Goumard}, {Leroy},
  {Martineau-Huynh}, {Lohse}, {Marcowith}, {Masterson}, {McComb}, {de Naurois},
  {Nolan}, {Noutsos}, {Orford}, {Osborne}, {Ouchrif}, {Panter}, {Pelletier},
  {Pita}, {P{\"u}hlhofer}, {Punch}, {Raubenheimer}, {Raue}, {Raux}, {Rayner},
  {Redondo}, {Reimer}, {Reimer}, {Ripken}, {Rob}, {Rolland}, {Rowell},
  {Sahakian}, {Saug{\'e}}, {Schlenker}, {Schlickeiser}, {Schuster}, {Schwanke},
  {Siewert}, {Skj{\ae}raasen}, {Sol}, {Steenkamp}, {Stegmann}, {Tavernet},
  {Terrier}, {Th{\'e}oret}, {Tluczykont}, {Vasileiadis}, {Venter}, {Vincent},
  {V{\"o}lk}, \& {Wagner}}]{2005A&A...442....1A}
{Aharonian}, F., {Akhperjanian}, A.~G., {Aye}, K.~M., {et~al.} 2005, \aap, 442,
  1, \dodoi{10.1051/0004-6361:20052983}

\bibitem[{{Ball} {et~al.}(1999){Ball}, {Melatos}, {Johnston}, \& {Skj{\ae}
  Raasen}}]{1999ApJ...514L..39B}
{Ball}, L., {Melatos}, A., {Johnston}, S., \& {Skj{\ae} Raasen}, O. 1999,
  \apjl, 514, L39, \dodoi{10.1086/311928}

\bibitem[{{Carciofi} \& {Bjorkman}(2006)}]{2006ApJ...639.1081C}
{Carciofi}, A.~C., \& {Bjorkman}, J.~E. 2006, \apj, 639, 1081,
  \dodoi{10.1086/499483}

\bibitem[{{Chang} {et~al.}(2021){Chang}, {Zhang}, {Chen}, {Ji}, {Kong}, \&
  {Wang}}]{2021Univ....7..472C}
{Chang}, Z., {Zhang}, S., {Chen}, Y.-P., {et~al.} 2021, Universe, 7, 472,
  \dodoi{10.3390/universe7120472}

\bibitem[{{Chen} {et~al.}(2019){Chen}, {Takata}, {Yi}, {Yu}, \&
  {Cheng}}]{2019A&A...627A..87C}
{Chen}, A.~M., {Takata}, J., {Yi}, S.~X., {Yu}, Y.~W., \& {Cheng}, K.~S. 2019,
  \aap, 627, A87, \dodoi{10.1051/0004-6361/201935166}

\bibitem[{{Chernyakova} {et~al.}(2020){Chernyakova}, {Malyshev}, {Mc Keague},
  {van Soelen}, {Marais}, {Martin-Carrillo}, \& {Murphy}}]{2020MNRAS.497..648C}
{Chernyakova}, M., {Malyshev}, D., {Mc Keague}, S., {et~al.} 2020, \mnras, 497,
  648, \dodoi{10.1093/mnras/staa1876}

\bibitem[{{Chernyakova} {et~al.}(2024{\natexlab{a}}){Chernyakova}, {Malyshev},
  {van Soelen}, {Mc Keague}, {O'Sullivan}, \& {Buckley}}]{2024MNRAS.528.5231C}
{Chernyakova}, M., {Malyshev}, D., {van Soelen}, B., {et~al.}
  2024{\natexlab{a}}, \mnras, 528, 5231, \dodoi{10.1093/mnras/stae265}

\bibitem[{{Chernyakova} {et~al.}(2006){Chernyakova}, {Neronov}, {Lutovinov},
  {Rodriguez}, \& {Johnston}}]{2006MNRAS.367.1201C}
{Chernyakova}, M., {Neronov}, A., {Lutovinov}, A., {Rodriguez}, J., \&
  {Johnston}, S. 2006, \mnras, 367, 1201,
  \dodoi{10.1111/j.1365-2966.2005.10039.x}

\bibitem[{{Chernyakova} {et~al.}(2021){Chernyakova}, {Malyshev}, {van Soelen},
  {O'Sullivan}, {Sobey}, {Tsygankov}, {Mc Keague}, {Green}, {Kirwan},
  {Santangelo}, {P{\"u}hlhofer}, \& {Monageng}}]{2021Univ....7..242C}
{Chernyakova}, M., {Malyshev}, D., {van Soelen}, B., {et~al.} 2021, Universe,
  7, 242, \dodoi{10.3390/universe7070242}

\bibitem[{{Chernyakova} {et~al.}(2024{\natexlab{b}}){Chernyakova}, {Malyshev},
  {van Soelen}, {Finn Galagher}, {Matchett}, {Russell}, {van den Eijnden},
  {Lower}, {Johnston}, {Tsygankov}, {Salganik}, \&
  {Shebalkova}}]{2024arXiv241102128C}
---. 2024{\natexlab{b}}, arXiv e-prints, arXiv:2411.02128,
  \dodoi{10.48550/arXiv.2411.02128}

\bibitem[{{Connors} {et~al.}(2002){Connors}, {Johnston}, {Manchester}, \&
  {McConnell}}]{2002MNRAS.336.1201C}
{Connors}, T.~W., {Johnston}, S., {Manchester}, R.~N., \& {McConnell}, D. 2002,
  \mnras, 336, 1201, \dodoi{10.1046/j.1365-8711.2002.05850.x}

\bibitem[{{Dubus}(2013)}]{2013A&ARv..21...64D}
{Dubus}, G. 2013, \aapr, 21, 64, \dodoi{10.1007/s00159-013-0064-5}

\bibitem[{{Fujita} {et~al.}(2019){Fujita}, {Kawachi}, {Akahori}, {Nagai}, \&
  {Yamaguchi}}]{2019PASJ...71L...3F}
{Fujita}, Y., {Kawachi}, A., {Akahori}, T., {Nagai}, H., \& {Yamaguchi}, M.
  2019, \pasj, 71, L3 (Papar I), \dodoi{10.1093/pasj/psz085}

\bibitem[{{Fujita} {et~al.}(2020){Fujita}, {Nagai}, {Akahori}, {Kawachi}, \&
  {Okazaki}}]{2020PASJ...72L...9F}
{Fujita}, Y., {Nagai}, H., {Akahori}, T., {Kawachi}, A., \& {Okazaki}, A.~T.
  2020, \pasj, 72, L9 (Papar II), \dodoi{10.1093/pasj/psaa067}

\bibitem[{{H.~E.~S.~S. Collaboration} {et~al.}(2024){H.~E.~S.~S.
  Collaboration}, {Aharonian}, {Ait Benkhali}, {Aschersleben}, {Ashkar},
  {Backes}, {Barbosa Martins}, {Batzofin}, {Becherini}, {Berge},
  {Bernl{\"o}hr}, {B{\"o}ttcher}, {Boisson}, {Bolmont}, {de Bony de Lavergne},
  {Borowska}, {Bouyahiaoui}, {Brose}, {Brown}, {Brun}, {Bruno}, {Bulik},
  {Burger-Scheidlin}, {Caroff}, {Casanova}, {Celic}, {Cerruti}, {Chand},
  {Chandra}, {Chen}, {Chibueze}, {Chibueze}, {Cotter}, {Damascene
  Mbarubucyeye}, {Devin}, {Djuvsland}, {Dmytriiev}, {Egberts}, {Einecke},
  {Ernenwein}, {Fontaine}, {Funk}, {Gabici}, {Gallant}, {Glawion},
  {Glicenstein}, {Goswami}, {Grolleron}, {Haerer}, {He{\ss}}, {Hofmann},
  {Holch}, {Holler}, {Huang}, {Jamrozy}, {Jankowsky}, {Joshi}, {Jung-Richardt},
  {Kasai}, {Katarzy{\'n}ski}, {Khangulyan}, {Khatoon}, {Kh{\'e}lifi},
  {Klu{\'z}niak}, {Komin}, {Kosack}, {Kostunin}, {Kundu}, {Lang}, {Le Stum},
  {Leitl}, {Lemi{\`e}re}, {Lemoine-Goumard}, {Lenain}, {Leuschner}, {Mackey},
  {Malyshev}, {Mart{\'\i}-Devesa}, {Marx}, {Mehta}, {Meintjes}, {Mitchell},
  {Moderski}, {Mohrmann}, {Montanari}, {Moulin}, {Murach}, {de Naurois},
  {Niemiec}, {Ohm}, {de Ona Wilhelmi}, {Ostrowski}, {Panny}, {Panter},
  {Parsons}, {Pensec}, {Peron}, {Prokhorov}, {P{\"u}hlhofer}, {Punch},
  {Quirrenbach}, {Regeard}, {Reimer}, {Reimer}, {Reis}, {Ren}, {Rieger},
  {Rudak}, {Ruiz-Velasco}, {Sahakian}, {Salzmann}, {Santangelo}, {Sasaki},
  {Sch{\"a}fer}, {Sch{\"u}ssler}, {Schutte}, {Shapopi}, {Spencer}, {Stawarz},
  {Steenkamp}, {Steinmassl}, {Steppa}, {Streil}, {Sushch}, {Takahashi},
  {Tanaka}, {Taylor}, {Terrier}, {Thorpe-Morgan}, {Tluczykont}, {Unbehaun},
  {van Eldik}, {van Soelen}, {Vecchi}, {Venter}, {Vink}, {Wach}, {Wagner},
  {Werner}, {Wierzcholska}, {Zacharias}, {Zdziarski}, {Zech}, \&
  {{\.Z}ywucka}}]{2024A&A...687A.219H}
{H.~E.~S.~S. Collaboration}, {Aharonian}, F., {Ait Benkhali}, F., {et~al.}
  2024, \aap, 687, A219, \dodoi{10.1051/0004-6361/202449612}

\bibitem[{{Hare} {et~al.}(2019){Hare}, {Kargaltsev}, {Pavlov}, \&
  {Beniamini}}]{2019ApJ...882...74H}
{Hare}, J., {Kargaltsev}, O., {Pavlov}, G., \& {Beniamini}, P. 2019, \apj, 882,
  74, \dodoi{10.3847/1538-4357/ab3648}

\bibitem[{{Johnston} {et~al.}(2005){Johnston}, {Ball}, {Wang}, \&
  {Manchester}}]{2005MNRAS.358.1069J}
{Johnston}, S., {Ball}, L., {Wang}, N., \& {Manchester}, R.~N. 2005, \mnras,
  358, 1069, \dodoi{10.1111/j.1365-2966.2005.08854.x}

\bibitem[{{Johnston} {et~al.}(1996){Johnston}, {Manchester}, {Lyne}, {D'Amico},
  {Bailes}, {Gaensler}, \& {Nicastro}}]{1996MNRAS.279.1026J}
{Johnston}, S., {Manchester}, R.~N., {Lyne}, A.~G., {et~al.} 1996, \mnras, 279,
  1026, \dodoi{10.1093/mnras/279.3.1026}

\bibitem[{{Johnston} {et~al.}(1999){Johnston}, {Manchester}, {McConnell}, \&
  {Campbell-Wilson}}]{1999MNRAS.302..277J}
{Johnston}, S., {Manchester}, R.~N., {McConnell}, D., \& {Campbell-Wilson}, D.
  1999, \mnras, 302, 277, \dodoi{10.1046/j.1365-8711.1999.02133.x}

\bibitem[{{Kawachi} {et~al.}(2021){Kawachi}, {Moritani}, {Okazaki}, {Yoshida},
  \& {Suzuki}}]{2021PASJ...73..545K}
{Kawachi}, A., {Moritani}, Y., {Okazaki}, A.~T., {Yoshida}, H., \& {Suzuki}, K.
  2021, \pasj, 73, 545, \dodoi{10.1093/pasj/psab019}

\bibitem[{{McMullin} {et~al.}(2007){McMullin}, {Waters}, {Schiebel}, {Young},
  \& {Golap}}]{2007ASPC..376..127M}
{McMullin}, J.~P., {Waters}, B., {Schiebel}, D., {Young}, W., \& {Golap}, K.
  2007, in Astronomical Society of the Pacific Conference Series, Vol. 376,
  Astronomical Data Analysis Software and Systems XVI, ed. R.~A. {Shaw},
  F.~{Hill}, \& D.~J. {Bell}, 127

\bibitem[{{Miller-Jones} {et~al.}(2018){Miller-Jones}, {Deller}, {Shannon},
  {Dodson}, {Mold{\'o}n}, {Rib{\'o}}, {Dubus}, {Johnston}, {Paredes}, {Ransom},
  \& {Tomsick}}]{2018MNRAS.479.4849M}
{Miller-Jones}, J.~C.~A., {Deller}, A.~T., {Shannon}, R.~M., {et~al.} 2018,
  \mnras, 479, 4849, \dodoi{10.1093/mnras/sty1775}

\bibitem[{{Negueruela} {et~al.}(2011){Negueruela}, {Rib{\'o}}, {Herrero},
  {Lorenzo}, {Khangulyan}, \& {Aharonian}}]{2011ApJ...732L..11N}
{Negueruela}, I., {Rib{\'o}}, M., {Herrero}, A., {et~al.} 2011, \apjl, 732,
  L11, \dodoi{10.1088/2041-8205/732/1/L11}

\bibitem[{{Okazaki} {et~al.}(2011){Okazaki}, {Nagataki}, {Naito}, {Kawachi},
  {Hayasaki}, {Owocki}, \& {Takata}}]{2011PASJ...63..893O}
{Okazaki}, A.~T., {Nagataki}, S., {Naito}, T., {et~al.} 2011, \pasj, 63, 893,
  \dodoi{10.1093/pasj/63.4.893}

\bibitem[{{Pavlov} {et~al.}(2011){Pavlov}, {Chang}, \&
  {Kargaltsev}}]{2011ApJ...730....2P}
{Pavlov}, G.~G., {Chang}, C., \& {Kargaltsev}, O. 2011, \apj, 730, 2,
  \dodoi{10.1088/0004-637X/730/1/2}

\bibitem[{{Rivinius} {et~al.}(2013){Rivinius}, {Carciofi}, \&
  {Martayan}}]{2013A&ARv..21...69R}
{Rivinius}, T., {Carciofi}, A.~C., \& {Martayan}, C. 2013, \aapr, 21, 69,
  \dodoi{10.1007/s00159-013-0069-0}

\bibitem[{{Shannon} {et~al.}(2014){Shannon}, {Johnston}, \&
  {Manchester}}]{2014MNRAS.437.3255S}
{Shannon}, R.~M., {Johnston}, S., \& {Manchester}, R.~N. 2014, \mnras, 437,
  3255, \dodoi{10.1093/mnras/stt2123}

\bibitem[{{Takata} {et~al.}(2012){Takata}, {Okazaki}, {Nagataki}, {Naito},
  {Kawachi}, {Lee}, {Mori}, {Hayasaki}, {Yamaguchi}, \&
  {Owocki}}]{2012ApJ...750...70T}
{Takata}, J., {Okazaki}, A.~T., {Nagataki}, S., {et~al.} 2012, \apj, 750, 70,
  \dodoi{10.1088/0004-637X/750/1/70}

\bibitem[{{Tavani} \& {Arons}(1997)}]{1997ApJ...477..439T}
{Tavani}, M., \& {Arons}, J. 1997, \apj, 477, 439, \dodoi{10.1086/303676}

\bibitem[{{Uchiyama} {et~al.}(2009){Uchiyama}, {Tanaka}, {Takahashi}, {Mori},
  \& {Nakazawa}}]{2009ApJ...698..911U}
{Uchiyama}, Y., {Tanaka}, T., {Takahashi}, T., {Mori}, K., \& {Nakazawa}, K.
  2009, \apj, 698, 911, \dodoi{10.1088/0004-637X/698/1/911}

\bibitem[{{van Soelen} \& {Meintjes}(2011)}]{2011MNRAS.412.1721V}
{van Soelen}, B., \& {Meintjes}, P.~J. 2011, \mnras, 412, 1721,
  \dodoi{10.1111/j.1365-2966.2010.18008.x}

\bibitem[{{van Soelen} {et~al.}(2012){van Soelen}, {Meintjes}, {Odendaal}, \&
  {Townsend}}]{2012MNRAS.426.3135V}
{van Soelen}, B., {Meintjes}, P.~J., {Odendaal}, A., \& {Townsend}, L.~J. 2012,
  \mnras, 426, 3135, \dodoi{10.1111/j.1365-2966.2012.21870.x}

\bibitem[{{van Soelen} {et~al.}(2016){van Soelen}, {V{\"a}is{\"a}nen},
  {Odendaal}, {Klindt}, {Sushch}, \& {Meintjes}}]{2016MNRAS.455.3674V}
{van Soelen}, B., {V{\"a}is{\"a}nen}, P., {Odendaal}, A., {et~al.} 2016,
  \mnras, 455, 3674, \dodoi{10.1093/mnras/stv2576}

\end{thebibliography}
\bibliographystyle{aasjournal}

\end{document}